\documentclass[aps,prl,twocolumn,superscriptaddress,floatfix,longbibliography]{revtex4-2}
\usepackage[colorlinks,bookmarks=false,citecolor=blue,linkcolor=red,urlcolor=blue]{hyperref}
\usepackage{epsfig}
\usepackage{tikz}
\usepackage{graphicx,comment}
\usepackage{enumitem,}
\setlist[itemize]{align=parleft,left=0pt..1em}

\usepackage{dcolumn}
\usepackage{bm}

\usepackage{amsmath,amssymb}
\usepackage{color,soul}

\usepackage{bbold}

\newcommand{\bea}{\begin{eqnarray}}          
	\newcommand{\eea}{\end{eqnarray}}

\allowdisplaybreaks

\begin{document}
	
	\title{Quantum loops in the 1T transition metal dichalcogenides}
	\author{Ashland Knowles}
	\email{gk23dp@brocku.ca}
	\affiliation{Department of Physics, Brock University, St. Catharines, Ontario L2S 3A1, Canada}
	\author{G. Baskaran}
	\email{baskaran@imsc.res.in}
	\affiliation{The Institute of Mathematical Sciences, CIT Campus, Chennai 600 113, India}
	\affiliation{Department of Physics, Indian Institute of Technology Madras, Chennai 600036, India}
	\affiliation{Perimeter Institute for Theoretical Physics, Waterloo, ON N2L 2Y5, Canada}
	\author{R. Ganesh}
	\email{r.ganesh@brocku.ca}
	\affiliation{Department of Physics, Brock University, St. Catharines, Ontario L2S 3A1, Canada}
	
	\date{\today}
	
	\begin{abstract}
		Loop arrangements and their quantum superpositions describe several interesting many-particle states. We propose that they also describe bonding in a class of transition metal dichalcogenides. We present an effective quantum loop model for monolayers with 1T structure and a d$^2$ valence electron configuration: materials of the form MX$_2$ (M = Mo, W and X=S, Se, Te) and AM$'$Y$_2$ (A = Li, Na; M$'$ = V, Nb and Y = O, S, Se). Their t$_{2g}$ orbitals exhibit strongly directional overlaps between neighbouring atoms, favouring the formation of valence bonds.
		A transition metal atom forms two valence bonds, each with one of its neighbours. When connected, these bonds form loops that cover the triangular lattice. 
		We construct a minimal Rokhsar-Kivelson-like model with resonance processes that cut and reconnect loops that run in proximity. The resulting dynamics is more constrained than in traditional quantum dimer models, with a `bending' constraint that arises from orbital structure. In the resulting phase diagram, we find phases that resemble distorted phases seen in materials, viz., the 1T$'$ and trimerized phases. As a testable prediction, we propose that a single d$^1$ or d$^3$ impurity will terminate a loop and give rise to a long-ranged texture. For example, a Ti/Cr defect in LiVO$_2$ will produce one or more domain walls that propagate outward from the impurity. We discuss the possibility of a loop liquid phase that can emerge in these materials.
	\end{abstract}
	
	\keywords{}
	\maketitle  
	
	\noindent{\color{blue} \it{Introduction}}---~
	Loops and loop-superpositions appear in many exciting contexts. For example, Kitaev's toric code has a spin-liquid ground state, arising from a quantum superposition of all possible ways of drawing loops on the square lattice\cite{Kitaev2003,Trebst2007}. A second example is spin-ice where each allowed ice state can be viewed as a loop-covering of the diamond lattice\cite{Jaubert2011,Gingras2014}. 
	Quantum loop models (QLMs) offer a paradigm to study such systems\cite{Ran2023,Ran2024}. They describe dynamics within the set of all loop-coverings of the underlying lattice. Each loop covering is assigned a potential energy arising from proximity between loop segments. Dynamics arises from tunnelling processes where proximate loops are cut and crossed.  
	Here, we motivate a QLM on physical grounds with a view to describe bonding in a class of transition metal dichalcogenides (TMDs). 
	
	The building block of TMDs is the  
	X-M-X trilayer, where M and X are transition metal and chalcogenide atoms\cite{Manzeli2017}. Each layer forms a triangular lattice. When these three layers are ABC-stacked, the resulting structure is denoted as 1T and is prone to distortions. A prominent example is the 1T$'$ structure which exhibits superconductivity\cite{Qi2016,Guo2017,Fang2018} and topological bands\cite{Qian2014,Shi2019}. We build a model for bonding in 1T structures with a view to describing distortions.

	\noindent{\color{blue} \it{Bonding considerations in d$^2$ 1T-TMDs}}---~
	As chalcogens are highly electronegative, they absorb two electrons to form X$^{2-}$. In MX$_2$ materials where M =Mo or W, this leaves two electrons in the valence d shell. Similarly, in AM$'$X$_2$ where A=Li, Na, and M$'$=V, Nb, the M$'$ atom is left in a d$^2$ configuration. This allows for each transition metal atom to form two bonds.
	
\begin{figure*}
		\includegraphics[width=2\columnwidth]{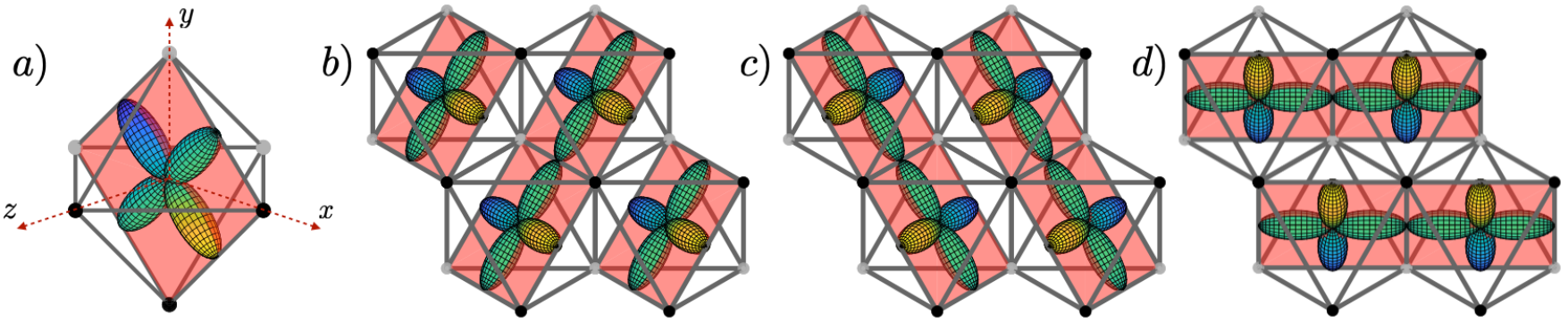}
		\caption{a) Octahedral coordination in the 1T structure: the transition metal atom surrounded by six chalcogen atoms that are located at $\pm \hat{x}$, $\pm \hat{y}$ and $\pm \hat{z}$. The figure shows a $d_{xy}$ orbital, lying in the $xy$ plane. Overlaps of t$_{2g}$ orbitals on nearest neighbour bonds: (b) $xy$ orbitals have strong overlaps along $\pm\{ \hat{x}+\hat{y}\}$ bonds, (c) $yz$ orbitals along $\pm\{ \hat{y}+\hat{z}\}$ bonds and (d) $zx$ orbitals along $\pm\{ \hat{z}+\hat{x}\}$ bonds.}
		\label{fig.overlaps}
	\end{figure*}

	In the 1T crystal structure, the octahedral crystal field splits the $d$ orbitals into t$_{2g}$ and e$_g$ levels\cite{Pavarini}. In a d$^2$ configuration, the two electrons occupy the lower-lying t$_{2g}$ orbitals. Crucially, in the undistorted 1T structure, the three t${_{2g}}$ orbitals show strong directionality in overlaps between neighbouring atoms. To see this, we first note that $xy$, $yz$ and $zx$ orbitals are defined with respect to a local coordinate system at each transition metal atom. The $x$, $y$ and $z$ axes point towards chalcogen atoms that lie at the vertices of an octahedron as shown in Fig.~\ref{fig.overlaps}(a). The six nearest neighbour vectors on the triangular lattice lie along $\pm (\hat{x} + \hat{y})$, $\pm (\hat{y}+\hat{z})$ and $\pm ( \hat{z} +\hat{x} )$. Along the $\pm (\hat{x}+\hat{y})$ directions, neighbouring atoms share the same $xy$ plane. Two such neighbours have a strong overlap between their d$_{xy}$ orbitals, and a zero overlap between any other pair of t$_{2g}$ orbitals. Similarly, the d$_{zx}$ and d$_{yz}$ orbitals have strong overlaps along $\pm (\hat{z}+\hat{x})$ and $\pm (\hat{y}+\hat{z})$ bonds respectively, as shown in Fig.~\ref{fig.overlaps}(b-d). 
	
	The directional nature of t$_{2g}$ overlaps favours the formation of valence bonds between nearest-neighbours of the M lattice. Furthermore, it imbues each such bond with sharp orbital character. 
	We gather these considerations into an effective quantum loop model below.

	\noindent{\color{blue} \it{Quantum loop model: Hilbert space}}---~We define the elements of our Hilbert space using two rules:
	\begin{itemize}[leftmargin=*]
		\item Dimers (valence bonds) are placed on nearest neighbour bonds of a triangular lattice, with two dimers touching each site. 
	\end{itemize}
	
	With two dimers at each site, we naturally form loops by tracing connected dimers. As a result, each allowed configuration can be viewed as a loop covering of the triangular lattice. 
	
	\begin{itemize}[leftmargin=*] 
		\item Two dimers that touch a site cannot be parallel.  
	\end{itemize}
	
	This rule arises from the orbital character of valence bonds. Suppose a certain site has two parallel bonds, say along $+ (\hat{x} + \hat{y})$ and $- (\hat{x} + \hat{y})$. As both bonds involve d$_{xy}$ orbitals, this site must have two electrons residing in its d$_{xy}$ orbital. This imposes a high energy cost due to intra-orbital Coulomb repulsion. This can be avoided by forcing the bonds to `bend', i.e., by demanding that bonds that touch at a site cannot be parallel.

\begin{figure}
		\includegraphics[width=0.8\columnwidth]{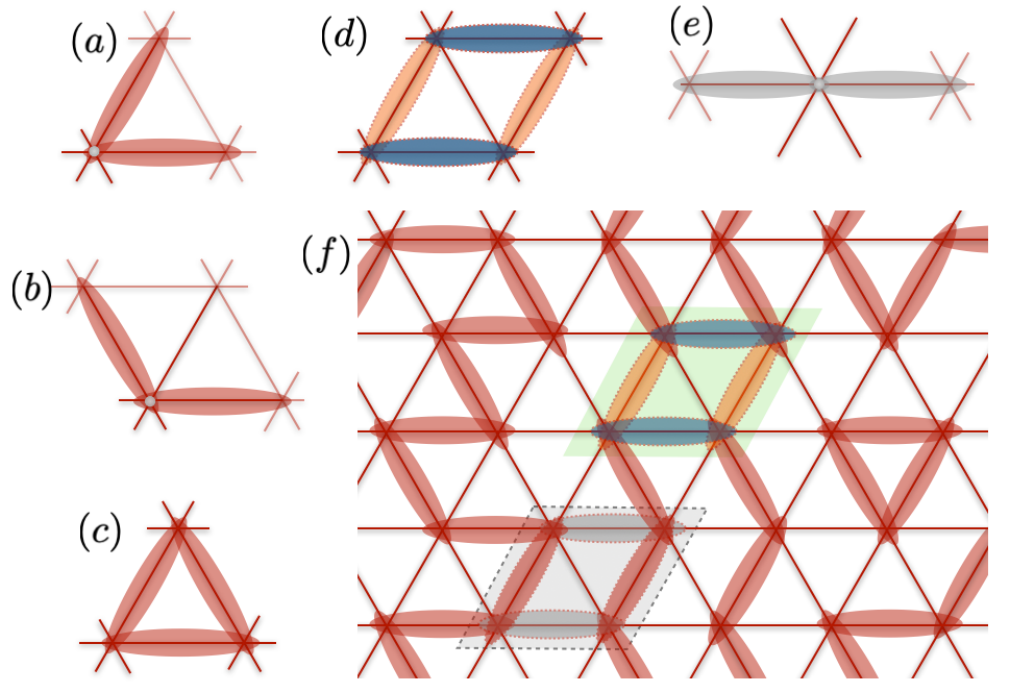}
		\caption{Elements of the QLM. Two dimers touch at each site, forming an acute angle (a) or an obtuse angle (b). Dimers connect to form loops, the shortest possible loop being a triangle (c). The simplest dynamical process operates on rhombus-shaped plaquettes, moving a pair of dimers from one pair of opposite edges to the other pair (d). Due to the bending constraint, parallel dimers \textit{cannot} touch at a site as shown in (e). A sample loop configuration is shown in (f). A possible cut-and-reconnect operation is shown in the top shaded region. A forbidden operation is shown in the bottom shaded region -- this operation will produce parallel dimers.    }
		\label{fig.qlm}
	\end{figure}

	\noindent{\color{blue} \it{Quantum loop model: Hamiltonian}}---~We now define a minimal Hamiltonian à la Rokhsar-Kivelson\cite{Rokhsar1988}. We first define potential energy with two contributions:
	\begin{enumerate}[label=(\roman*),wide, labelwidth=!, labelindent=0pt]
		\item The two dimers that touch at a site must either form an acute (60$^\circ$) or an obtuse angle (120$^\circ$) as shown in Fig.~\ref{fig.qlm}(a,b). We represent the relative energy between these local configurations by a single parameter. We assign an energy cost $V$ to each acute angle and zero to each obtuse angle. 
		\item The shortest possible loops are triangles on elementary plaquettes, with three valence bonds in close proximity as shown in Fig.~\ref{fig.qlm}(c). We assign an energy $V'$ to each triangle.    
	\end{enumerate}
	These energy contributions encode proximity of bonds. For example, electrons come closer when bonds form an acute angle rather than an obtuse angle. This can impose an energy cost due to Coulomb repulsion, corresponding to a positive value of $V$. In a triangle loop, we have three bonds that are very close to one another. This could give rise to a local distortion with the three vertices coming closer to one another, with an associated energy cost $V'$. We treat $V$ and $V'$ as independent parameters that can each take positive or negative values.

	We next consider dynamical processes, contributing to kinetic energy. We keep the simplest term, i.e., the smallest possible rearrangement of dimers that does not alter the environment. This involves a rhombus-shaped plaquette with dimers on a pair of opposite edges. These dimers can be moved to the other pair \textit{if} this does not create parallel bonds. 
	Examples of allowed and forbidden moves are shown in Fig.~\ref{fig.qlm}(f). Allowed processes can be viewed as cutting and crossing two loops.
	
	With these terms, we write a Hamiltonian 
	\begin{align}
		\nonumber \hat{H} = &\sum_{i=1}^{N}\left[-t \sum_{j=1}^{3} 
		\Big\{ \big\vert  \begin{gathered}\includegraphics[width=0.25in]{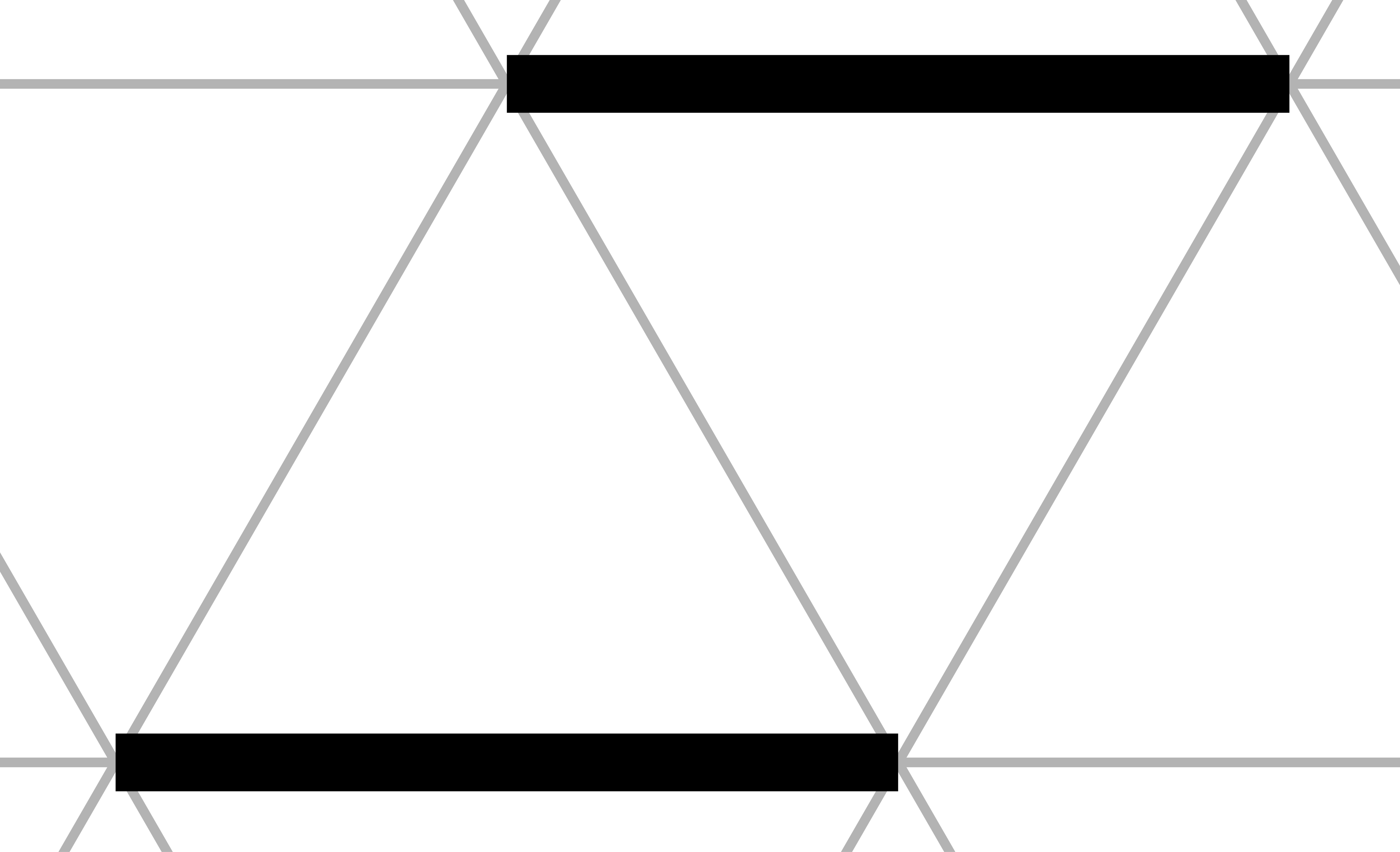} \end{gathered}
		\big\rangle \big\langle \begin{gathered} \includegraphics[width=0.25in]{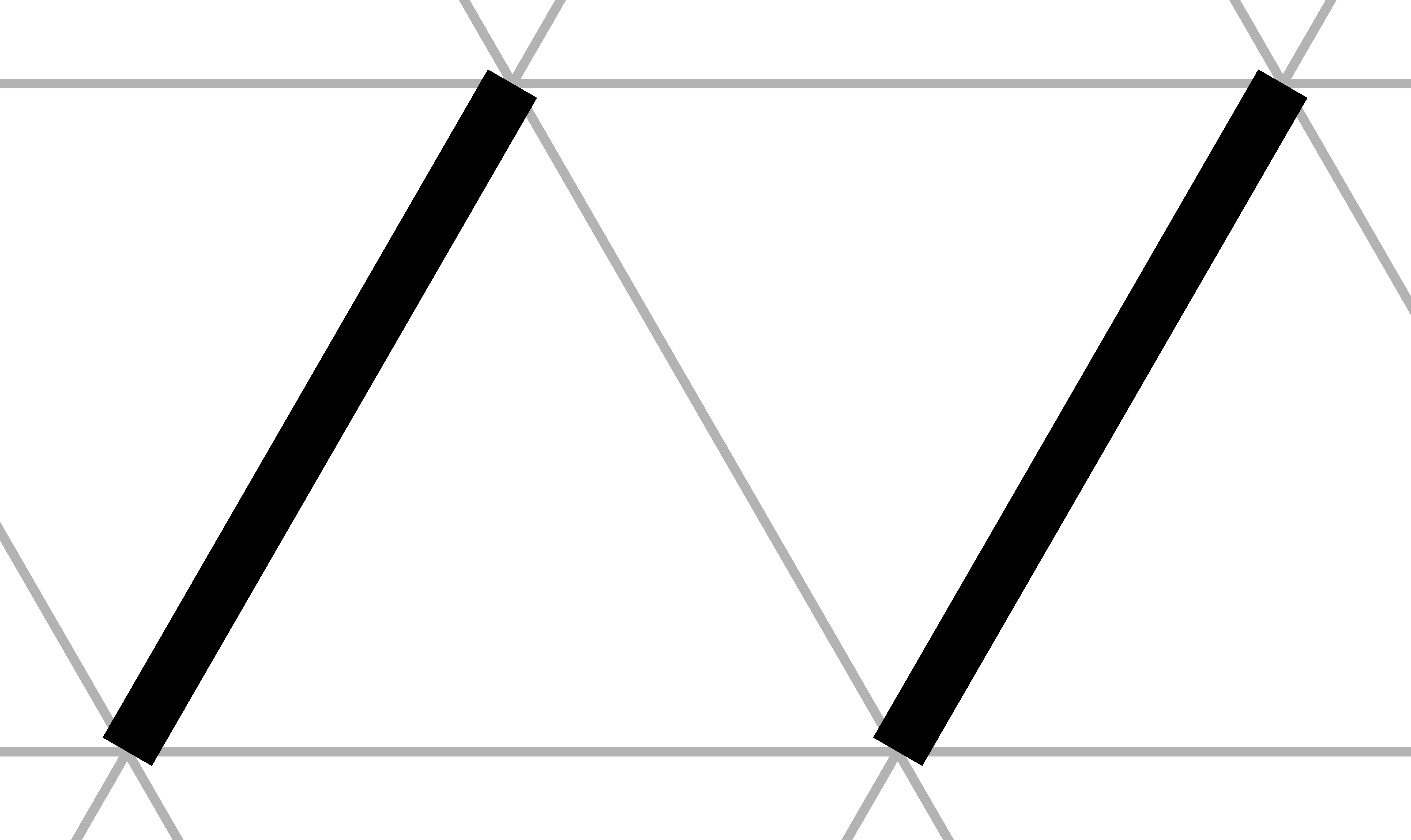}\end{gathered} \big\vert 
		+\big\vert  \begin{gathered}\includegraphics[width=0.25in]{KEF1A_new} \end{gathered}
		\big\rangle \big\langle \begin{gathered} \includegraphics[width=0.25in]{KEF1B_new}\end{gathered} \big\vert 
		\Big\}\right.\\
		&\left. +V \sum_{k=1}^{6} 
		\big\vert  \begin{gathered}\includegraphics[width=0.25in]{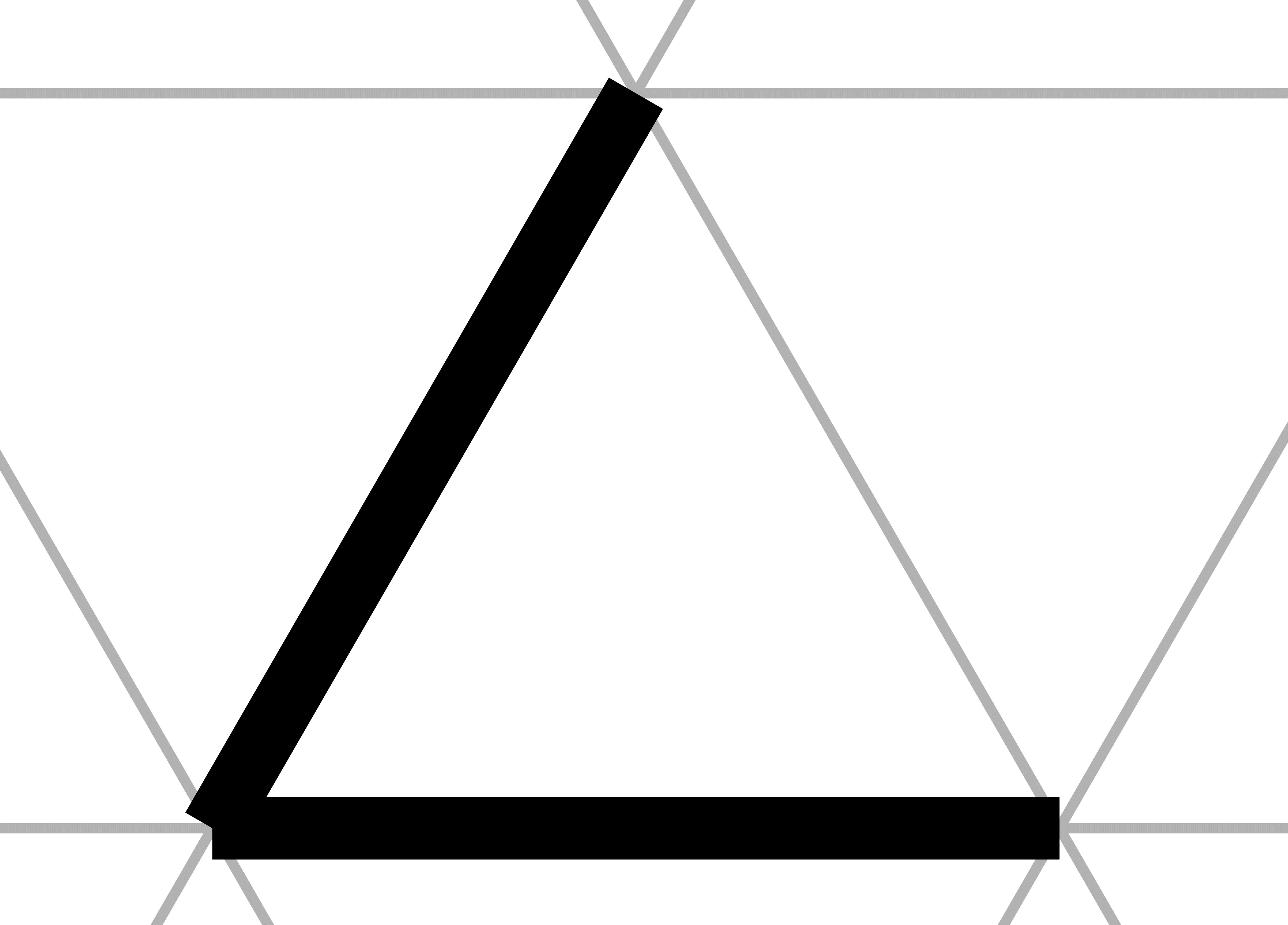} \end{gathered}
		\big\rangle \big\langle \begin{gathered} \includegraphics[width=0.25in]{PE1_new}\end{gathered} \big\vert 
		+V^{\prime}\sum_{l=1}^{2}\big\vert  \begin{gathered}\includegraphics[width=0.25in]{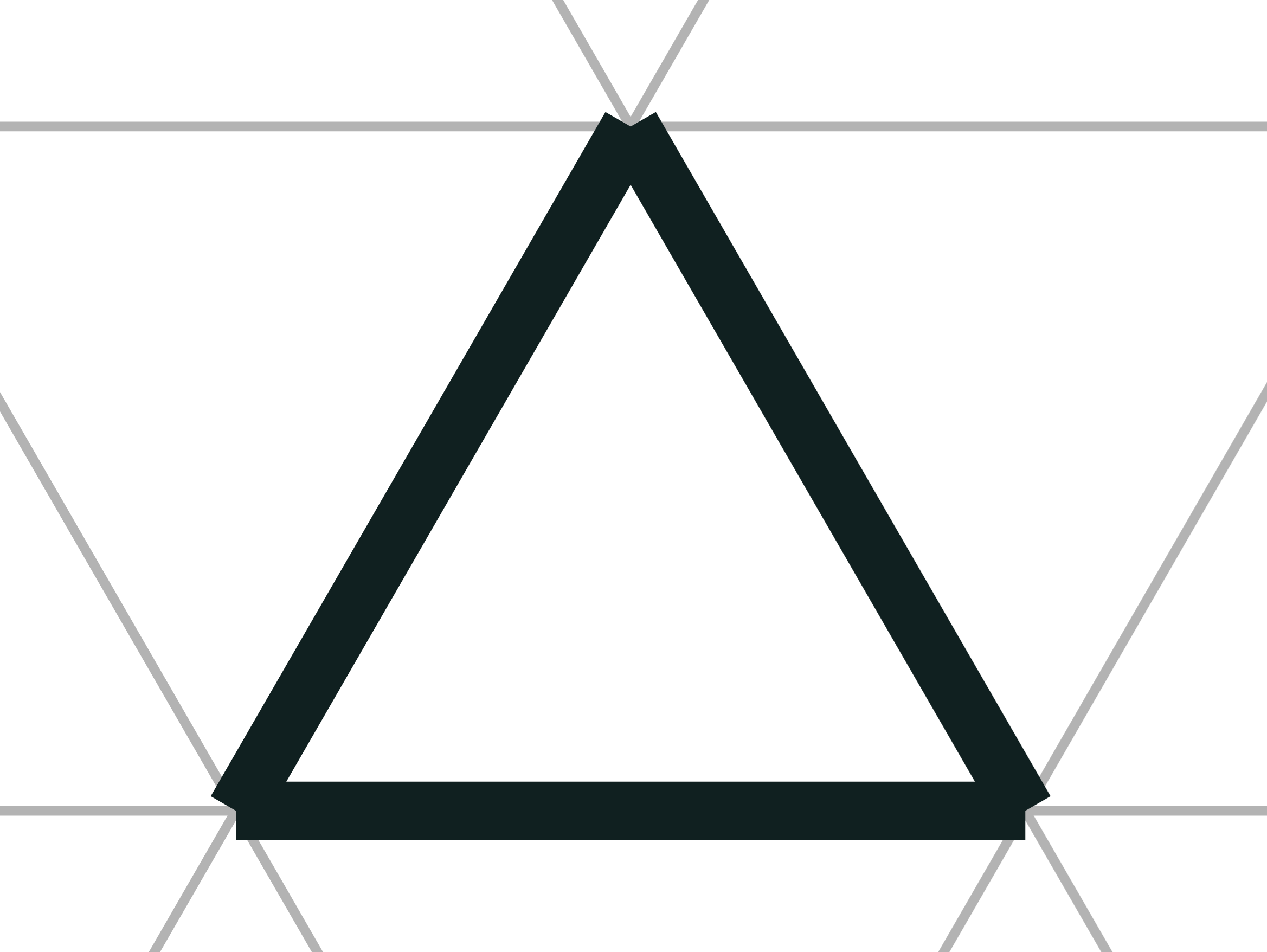} \end{gathered}
		\big\rangle \big\langle \begin{gathered} \includegraphics[width=0.25in]{Tri1}\end{gathered} \big\vert 
		\right].
		\label{eq.Hamiltonian}
	\end{align}
	The sum over $i$ runs over all sites of the triangular lattice. The $t$ terms involve a further sum over three rhombus orientations attached to each site. Similarly, the $V$ ($V'$) terms sum over six (two) possible orientations of acute angles (triangles) at site $i$.

	This Hamiltonian differs from previous studies that have adapted Rokhsar-Kivelson-like terms to a loop setting\cite{Ran2023,Ran2024}. There is a difference even at the level of the Hilbert space, due to the bending constraint at each site that arises from orbital character. This further constrains the kinetic energy term -- a flipping process is only allowed if the initial and final configurations satisfy the bending constraint. The second key difference is in the form of the potential energy terms. Previous studies have defined potential energy in direct analogy with the original Rokhsar-Kivelson model, associating an energy cost with each flippable plaquette. Here, we have defined potential energy on physical grounds motivated by the geometry and physics of TMDs.

	\noindent{\color{blue} \it{Numerical approach}}---~
	To solve for the ground state of Eq.~\ref{eq.Hamiltonian}, we first enumerate all loop coverings to construct the Hilbert space. This is a challenging task, with no known analytic solution.
	We use a numerical approach, working with finite clusters and periodic boundaries. 
	We use a recursive branching algorithm to find all loop coverings that satisfy the bending constraint. The number of configurations grows rapidly with system size: 42 for 3x3, 250 964 for 5x4, 2 720 400 for 6x4. For concreteness, we present results for the 6x4 cluster below. We expect the same qualitative behaviour with larger sizes. 
	
	We construct the Hamiltonian as a 2 720 400 $\times$ 2 720 400 matrix, with flipping processes entering in off-diagonal terms. The resulting matrix is sparse, with $\sim 10^{-4}\%$ of the entries being non-zero. We use a Krylov-space based approach to find the lowest-lying eigenstates. 
	
	\noindent{\color{blue} \it{Possible phases and energy estimates}}---~
	Before presenting the ground state phase diagram, we discuss some candidate phases and estimate their energies. The phases discussed below are depicted in Fig.~\ref{fig.states}(a-d).
	\begin{enumerate}[label=(\roman*),wide, labelwidth=!, labelindent=0pt]
		\item Acute stripe: This phase is dominated by a single configuration with stripes that form an acute angle at every site. Its potential energy can be estimated as $E_{PE}^{acute} \sim NV$, where $N$ is the number of sites in the cluster. This is a ground state candidate when $V$ takes large negative values.

		\begin{figure*}
			\includegraphics[width=2\columnwidth]{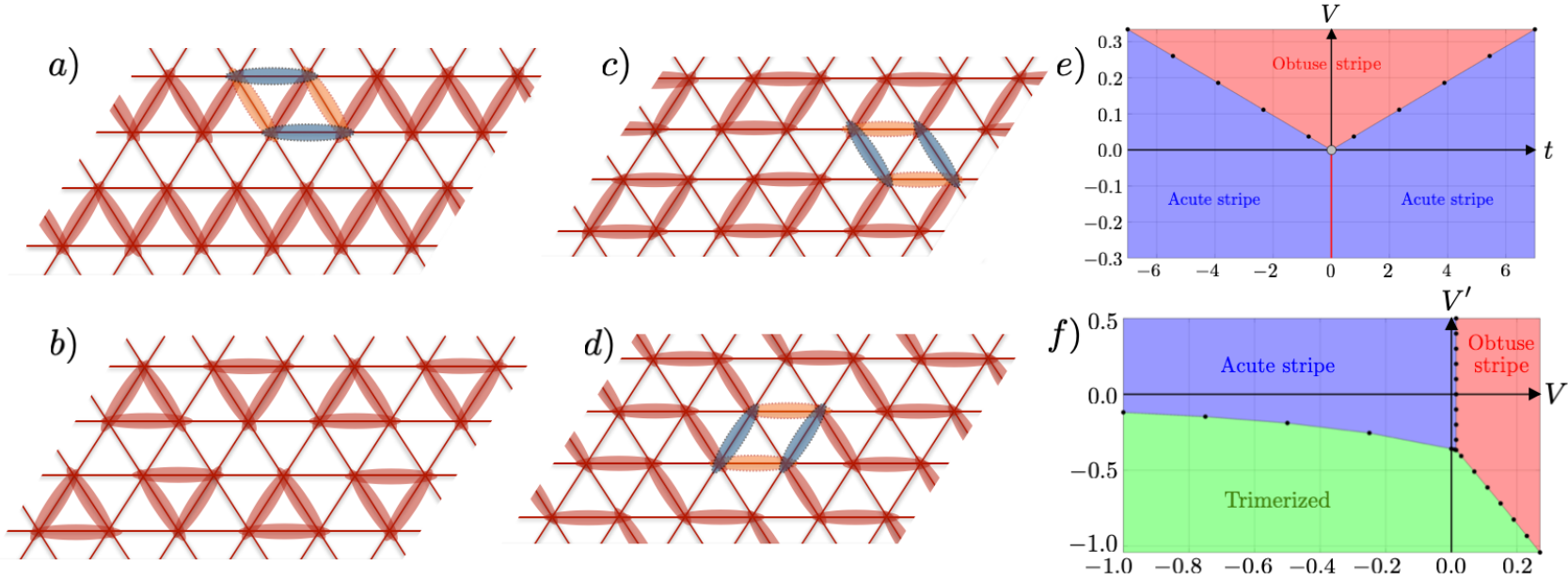}
			\caption{High symmetry configurations on a 6$\times$4 cluster: a) acute stripe, b) trimerized, c) rhombi and d) obtuse stripes. (a), (c) and (d) contain `flippable' plaquettes -- with one example shown in each. e) Slice of the phase diagram with $V'$ set to zero, varying $t$ and $V$. 
				The ($V<0$, $V'=0$, $t=0$) line separates two regions with the acute stripe phase. Along this line, we have 42 degenerate ground states, each of which has an acute angle at every site. The introduction of a small non-zero $t$ suffices to select the acute stripe phase on either side of this line. 
				f) Slice of the phase diagram with $t=0.3$ held fixed and while $(V,V')$ are varied. Apart from acute and obtuse stripes, a trimerized (triangle) phase is seen. 
			}
			\label{fig.states}
		\end{figure*}

		The acute stripe configuration is highly conducive to dynamics, as it allows for $N$ flips -- see Fig.~\ref{fig.states}(a). Each flip produces two obtuse angles, leading to an energy cost $\sim -2V$ (assuming $V<0$). In the limit of dominant potential energy ($\vert t\vert \ll \vert V\vert$, $V<0$), we estimate the energy of this phase to be $E_{acute}\sim NV - N\vert t\vert^2/2\vert V\vert$.
		
		\item Obtuse stripe: This phase is dominated by a single configuration with stripes that form an obtuse angle at every site. With neither acute angles nor triangles, potential energy vanishes. This state is a ground state candidate when $V$ takes large positive values. 
		
		The obtuse stripe configuration allows for $N$ flips. Each flip produces two acute angles, leading to an energy cost $\sim 2V$ (assuming $V>0$). When ($V>0$, $\vert t \vert\ll V$), we estimate the energy of this phase to be $E_{obtuse}\sim -  N\vert t\vert^2/2V$.
		
		\item Trimerized: Each site hosts an acute angle and is part of a triangle. This leads to a potential energy $E_{PE}^{trimer} \sim N(V +V'/3)$. This is a ground state candidate when $V,V'<0$. Kinetic energy terms are ineffective as there are no flippable plaquettes. Any flip will result in parallel dimers, violating the bending constraint.
		
		\item Rhomboid: With loops forming rhombi, we have $N/2$ acute and $N/2$ obtuse angles. We estimate potential energy to be $E_{PE}^{rhombi} \sim NV/2$. Flipping processes may fuse two rhombi, creating two obtuse angles in the process. 
	\end{enumerate}

	\noindent{\color{blue} \it{Phase diagram}}---~
	We consider the Hamiltonian in Eq.~{\ref{eq.Hamiltonian}} as a function of three parameters, $t$, $V$ and $V'$. Using exact diagonalization, we find the ground state as a linear superposition of all loop covers. To determine the nature of the ground state, we  examine (a) the loop cover(s) with the largest weight, and (b) dimer-dimer correlations in the ground state.
	We identify three phases:

	(i) Acute stripe: This phase appears for $V<0$, where acute angles are favoured.
	The acute stripe configuration maximizes the number of acute angles. However, there are many other configurations with the same number of acute angles, e.g., the trimerized phase. Among these states, the acute stripe configuration allows for the largest number of flipping processes. As a result, it is `selected' by kinetic energy as long as $t$ is non-zero. The ground state is a linear superposition of a large number of configurations that can be reached by flipping plaquettes starting from an acute stripe configuration. The precise distribution of weights varies with parameters. 
	
	We identify the acute stripe phase with the 1T$'$ phase\cite{Yin2021,Tang2021,Dai2024} seen in many materials. Bond correlations
	in the QLM ground state show a clear stripe-like pattern. This pattern is very similar to the variation of bond lengths seen in 1T$'$ materials -- see Supplemental Material\cite{supplementary}.
	In the QLM ground state, a stronger bond is indicated by a higher likelihood of dimer occupancy. In materials, a stronger bond is one with a shorter bond length.
	
	(ii) Obtuse stripe phase: This phase appears at positive values of $V$ where obtuse angles are favoured. Indeed, the obtuse stripe configuration maximizes the number of obtuse angles with all other configurations having fewer obtuse angles. They also allow for a high degree of dynamics with many flippable plaquettes. The ground state is a linear superposition with dominant weight from an obtuse stripe configuration.
	
	(iii) Trimerized phase: We find a single-configuration-state that contains an arrangement of triangle-loops. On the 6$\times$4 lattice, we find a trimerized phase with alternating `up' and `down' triangles as shown in Fig.~\ref{fig.states}(b). On other sizes, we find configurations with all-up or all-down -- see Supplemental Material\cite{supplementary}.  Trimerized phases do not allow for flipping processes within our model. Dynamics may appear in a more general model with longer-range rearrangements.

	A trimerized phase is known to appear in AVX$_2$, where A=Li, Na and X=O,S,Se. Previous studies have described trimerization as arising from orbital ordering\cite{PenPRL1997,PenPRB1997,Ezhov1998}. Our QLM places the trimerized phase within the broader context of loop model phases.

	\noindent{\color{blue} \it{Impurity textures}}---~
	Our QLM description is a real-space approach based on short-ranged valence bond formation. This can be contrasted with momentum-space-based studies, e.g., where distortions are viewed as exciton-like instabilities\cite{Monney2011,Kogar2017,Si2020} or as Peierls transitions\cite{Duerloo2014,Keum2015,Li2016}. To build a case for a real-space approach, we present a testable prediction. We consider a single impurity where a d$^1$ transition metal substitutes at the site of the d$^2$ metal atom. For example, we may have a Nb atom at a Mo site in 1T-MoS$_2$. A d$^1$ atom can only form a single valence bond. As a result, it cannot allow a loop to pass through. Rather, it serves as a loop-termination point. 
	
\begin{figure}
		\includegraphics[width=0.45\columnwidth]{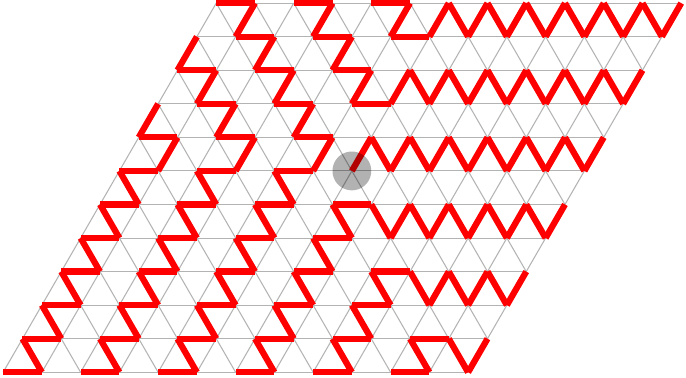}
		\includegraphics[width=0.45\columnwidth]{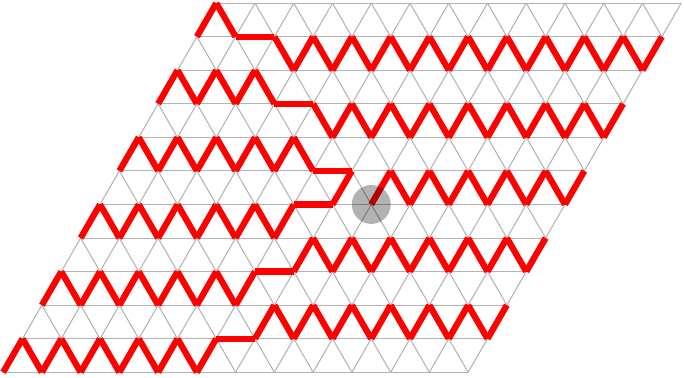}\\
		\vspace{0.1in}
		\includegraphics[width=0.45\columnwidth]{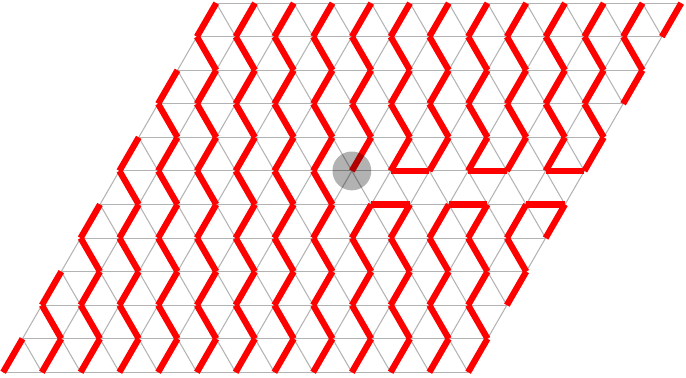}
		\includegraphics[width=0.45\columnwidth]{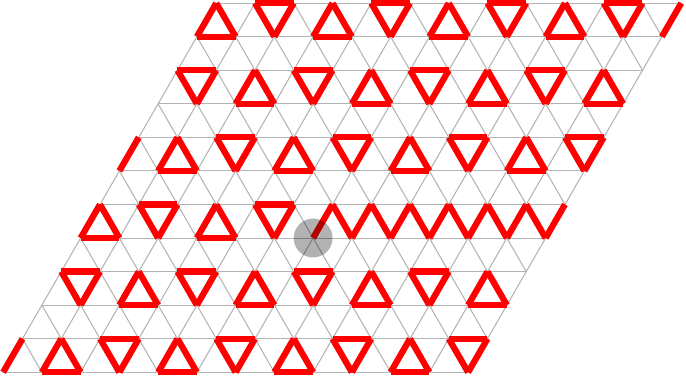}
		\caption{Textures generated by a single d$^1$ impurity. Figures show the neighbourhood of an impurity within a larger system. Configurations in the acute stripe (top left and top right), obtuse stripe (bottom left) and trimerized (bottom right) phases.    }
		\label{fig.impurities}
	\end{figure}

	Remarkably, a single loop-termination point disrupts the formation of a uniform phase and leads to a long-ranged texture. Fig.~\ref{fig.impurities} shows loop configurations in the presence of a single d$^1$ impurity. Fig.~\ref{fig.impurities}(top-left) and (top-right) depict a regime where acute stripes are favoured by the QLM Hamiltonian. We necessarily have multiple domains with symmetry-related stripe patterns. They are separated by domain walls emanating from the impurity. Such impurity-textures may appear in the 1T$'$ phase, generating domain walls that are visible to scanning probes. 
	Fig.~\ref{fig.impurities}(bottom-left) depicts an obtuse stripe regime, where a single domain wall emanates from the impurity. Fig.~\ref{fig.impurities}(bottom-right) shows a trimerized configuration, but with a stripe-like feature that emanates from the impurity. This texture may be realized by implanting a Ti atom at a V site in LiVO$_2$. Apart from the configurations shown in Fig.~\ref{fig.impurities}, there are many other possible textures that can be easily drawn. The common feature in all of them is that long-ranged domain walls emanate from the impurity. 
	
	As a further test, we may consider a d$^3$, rather than d$^1$, impurity. This creates a T-junction rather than a loop termination point. This leads to precisely the same distortion patterns as d$^1$, as shown in the Supplemental Material\cite{supplementary}.

	\noindent{\color{blue} \it{Discussion}}---~
	We have proposed a QLM description for 1T TMDs, where distortions are driven by short-ranged valence bond formation. Loops emerge from d$^2$ character and the directionality of t$_{2g}$ overlaps, with each loop containing a sequence of valence bonds. This picture is justified in a Mott insulator driven by strong intra-orbital Hubbard repulsion. If Hund's coupling is weak, we may identify each valence bond as a covalent bond or a spin-singlet wavefunction. If Hund's coupling is strong, two electrons on the same atom cannot form independent singlet bonds. Rather, electron spins will align to give rise to an effective spin-1 moment on the atom. Nevertheless, the QLM description can hold -- if each loop is interpreted as an emergent Haldane chain. This picture has been evoked to describe the trimerized phase of LiVO$_2$\cite{PenPRB1997}. Based on the same idea, loop models have been proposed on the square lattice in the context of LaFeAsO\cite{Baskaran2008} and more recently on the honeycomb lattice\cite{Savary2021}. 
	
	In MoX$_2$ and WX$_2$ (X=S,Se), the 1T$'$ structure is known to have a small electronic gap\cite{Keum2015,Chen2018}. However, the 1T structure is metallic\cite{Yin2021}. Naively, this goes against our premise of localized electrons. If these distortion-free materials are described by a QLM, they may arise from a putative loop-liquid state. This state may host mobile charged excitations, reminiscent of holons and doublons in the resonating valence bond (RVB) theory of high-temperature superconductivity. This is an exciting direction for future studies, motivating the search for a quantum loop liquid in analogy with quantum spin liquids. Notably, a loop liquid phase has been seen in quantum Monte Carlo simulations of a related loop model\cite{Ran2024}.

	\noindent{\color{blue} \it{Acknowledgments}}---~This work was supported by the Natural Sciences and Engineering Research Council of Canada through Discovery Grant 2022-05240. 
	Research at Perimeter Institute is supported in part by the Government of Canada through the Department of Innovation, Science and Economic Development Canada and by the Province of Ontario through the Ministry of Economic Development, Job, Creation and Trade.

	\bibliography{TMDloops}

	\newpage \clearpage
	
	\onecolumngrid 
	\setcounter{secnumdepth}{3}
	\renewcommand{\theequation}{S\arabic{equation}}
	\renewcommand{\thefigure}{S\arabic{figure}}
	\begin{center}
		\textbf{\large Supplemental Material for ``Quantum loops in the 1T transition metal dichalcogenides''}\\[.5cm]
		A. Knowles,$^1$ G. Baskaran,$^{2,3,4}$ and R. Ganesh$^1$\\[.4cm]
		{\itshape ${}^1$Department of Physics, Brock University, St. Catharines, Ontario L2S 3A1, Canada\\
			\itshape ${}^2$The Institute of Mathematical Sciences, CIT Campus, Chennai 600 113, India\\
			\itshape ${}^3$Department of Physics, Indian Institute of Technology Madras, Chennai 600036, India\\
			\itshape ${}^4$Perimeter Institute for Theoretical Physics, Waterloo, ON N2L 2Y5, Canada}\\
		
		(Dated: \today)\\[2cm]
	\end{center}

	\section{Recursive algorithm for listing loop configurations}
	We consider an $\ell_1 \times \ell_2$ triangular lattice with periodic boundaries, with $\ell_1$ and $\ell_2$ being the number of sites along the two primitive lattice vectors. We represent this lattice (graph) as an $\mathcal{N}\times \mathcal{N}$ matrix, where $\mathcal{N}=\ell_1\times\ell_2$. All entries in the matrix are set to zero initially. We seek to place non-zero entries (with value unity) to mark the presence of valence bonds. An entry of unity at row $i$ and column $j$ represents a valence bond from site $i$ to site $j$ -- this entry is only allowed if $i$ and $j$ are nearest neighbours. This matrix must be symmetric by definition -- when an entry is placed at $(i,j)$, we simultaneously place an entry at $(j,i)$ as well.  In the $i^{\mathrm{th}}$ row of the matrix, there are six allowed positions for unity -- corresponding to the six nearest neighbours of site $i$. However, due to the bending constraint, we cannot have valence bonds on opposite bonds simultaneously. This leads to 6$\times$4/2 = 12 choices for each row -- with six nearest-neighbour bonds at each site, we have 6 possibilities for the first valence bond; 4 for the second as to avoid parallel bonds and a factor of 2 to avoid double counting. 
	When placing entries on a given row, we check for consistency with all previously filled entries to ensure that (i) previously assigned bonds are retained and (ii) no parallel bonds are created. Formulating this process as a recursive algorithm, we generate all allowed loop coverings of the triangular lattice. As described in the main text, the number of loop coverings grows rapidly with system size. 
	
	The choice of system size ($\ell_1$ and $\ell_2$) constrains the allowed loop configurations. For example, an acute stripe configuration cannot be accommodated if both $\ell_1$ and $\ell_2$ are odd. In the main text, we present results for a 6$\times$4 lattice -- a choice that is capable of hosting multiple phases. Having $\ell_1 \neq \ell_2$ breaks rotational symmetry, e.g., with a lower energy for acute stripes that run along the longer direction. In the thermodynamic limit as well as in materials, all symmetry-related choices will be equally likely. 
	
	\section{Bond correlations}
	
\begin{figure}
		\includegraphics[width=0.8\columnwidth]{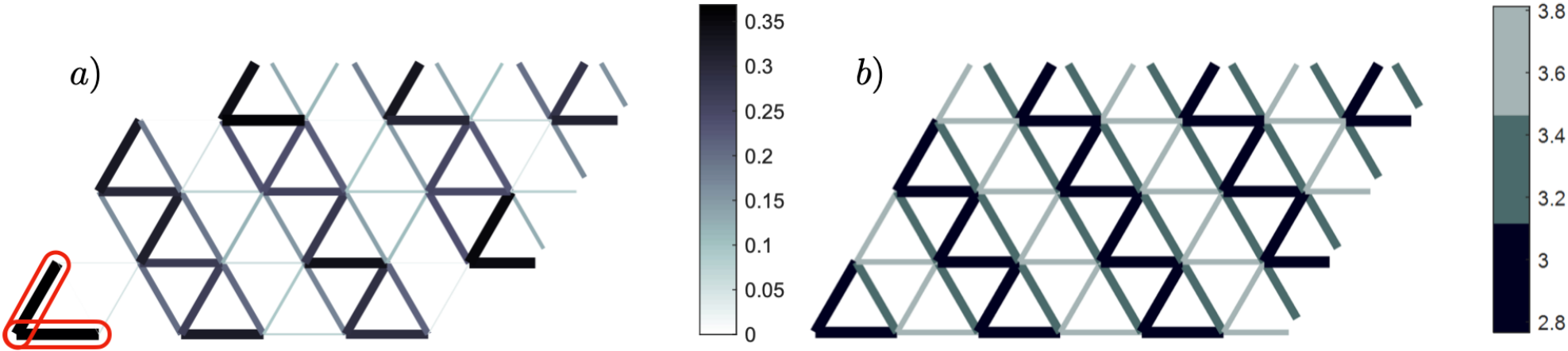}
		\caption{a) Dimer density in the acute stripe ground state of the QLM. The ground state is projected to only retain states where dimers are present at the two bonds at the bottom left. Bond colour and line width are proportional to the dimer probability.  
			b) Bond lengths in the MoS$_2$ using data from Ref.~\onlinecite{Qian2014}. Darker, thicker bonds have shorter bond lengths while lighter, thinner bonds are longer. }
		\label{fig.bondcorrs}
	\end{figure}

	We identify the acute stripe phase seen in the QLM with the 1T$'$ structure of the MX$_2$ family of materials. To bolster this assertion, we examine bond correlations in the QLM ground state. Fig.~\ref{fig.bondcorrs} shows dimer density after projecting to states that contain dimers at two reference bonds. The projection operation is carried out to separate a single stripe from its symmetry-equivalent partners (e.g., an iso-energetic phase where the stripes are translated in the `perpendicular' direction). The reference bonds can be viewed as pinning centres that fix the direction of stripe ordering in a real material. The resulting bond correlation plot shows a stripe-like pattern, but with varying dimer weights. This variation occurs due to quantum fluctuations -- due to contributions from sub-dominant loop coverings that are reachable from the dominant acute stripe configuration. This pattern is comparable to the bond lengths in MoS$_2$ as shown in the figure. A higher dimer density indicates a stronger bond, in turn, corresponding to a shorter bond length.

	\section{Trimerized phases}
\begin{figure}
		\includegraphics[width=0.3\columnwidth]{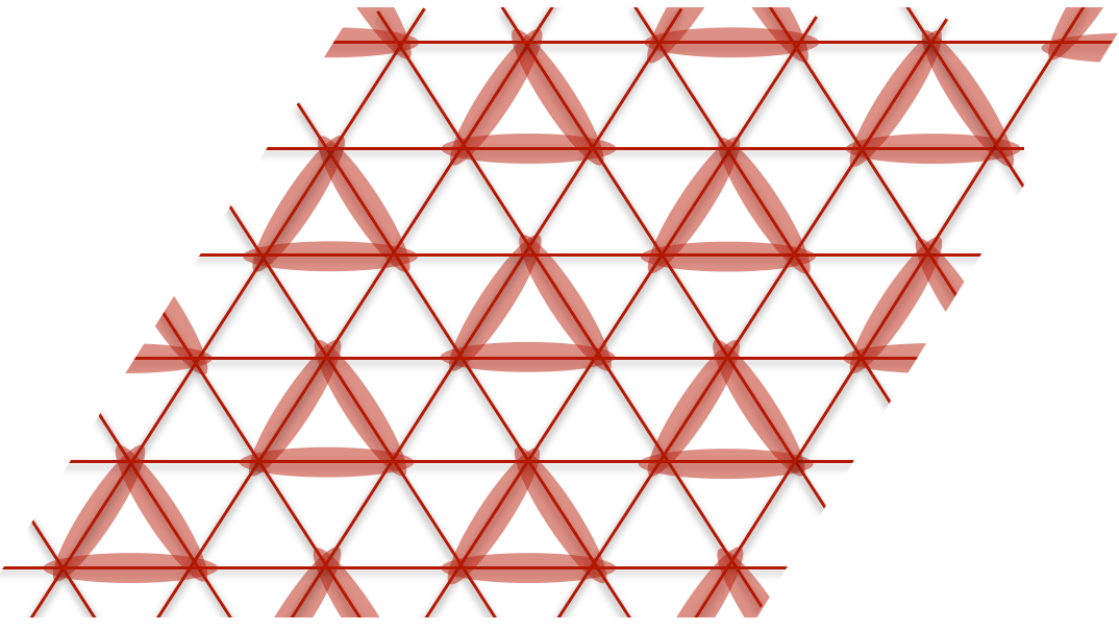}
		\caption{Trimerized configuration on a 6$\times$6 lattice. }
		\label{fig.uptriangles}
	\end{figure}

	In the main text, Fig.~\ref{fig.states}(b) shows a trimerized state where every site is part of a triangle-loop. The triangles alternate in orientation, with `up' triangles and `down' triangles next to one another. Fig.~\ref{fig.uptriangles} shows a 6$\times$6 lattice with a trimerized state where all triangles have the same orientation. Within our approach, the relative orientation is fixed by the system size -- the triangle pattern must respect the imposed periodicity. Any such trimerized state is immediately seen to be an eigenstate of our QLM Hamiltonian. There are no flippable plaquettes and therefore, no dynamics. The potential energy is the same (on a per site site basis) for any trimerized state. 
	
	In LiVO$_2$ and related materials, the trimerized phase has triangles with the same orientation -- closer to Fig.~\ref{fig.uptriangles} rather than Fig.~\ref{fig.states}(b) of the main text. A preference for parallel orientations can arise in a more general QLM Hamiltonian with longer-range potential energy terms.

	\section{\MakeLowercase{d}$^3$ impurity}
	
\begin{figure}
		\includegraphics[width=0.3\columnwidth]{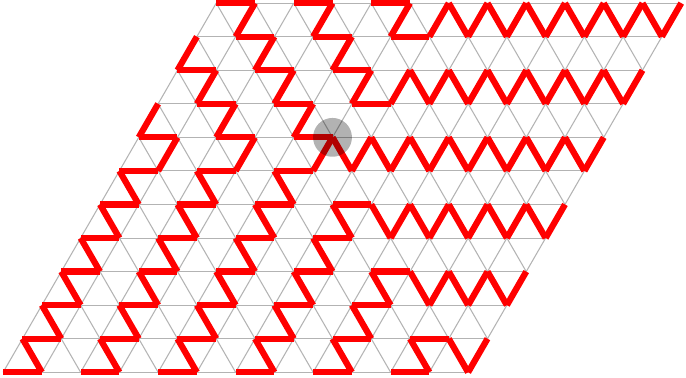}
		\includegraphics[width=0.3\columnwidth]{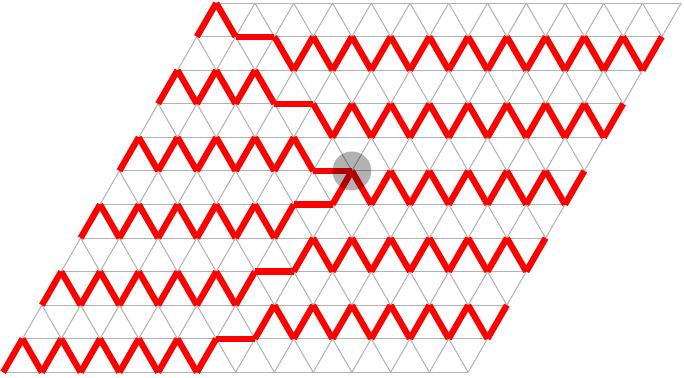}\\
		\vspace{0.1in}
		\includegraphics[width=0.3\columnwidth]{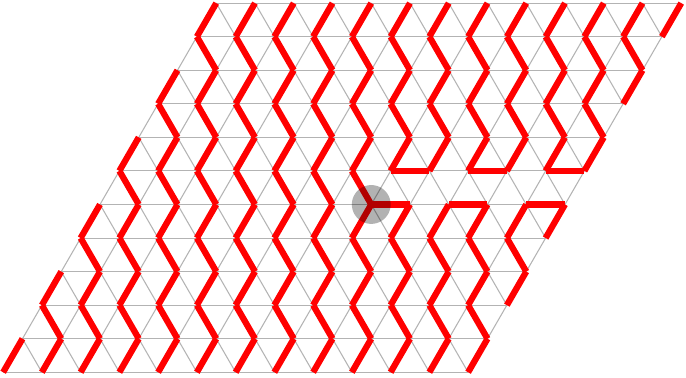}
		\includegraphics[width=0.3\columnwidth]{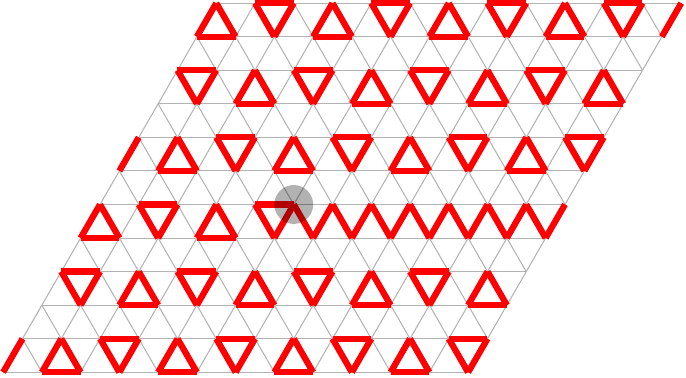}
		\caption{Textures generated by a single d$^3$ impurity. Configurations in the acute stripe (top left and top right), obtuse stripe (bottom left) and trimerized (bottom right) phases. These can be compared with Fig.~\ref{fig.impurities} in the main text.  }
		\label{fig.d3impurities}
	\end{figure}

	In the main text, we have discussed textures induced by a d$^1$ impurity as a testable prediction. We now argue that d$^3$ impurities can also be used with the same effect. For example, we may have a Re impurity substituting for W in 1T-WS$_2$ or a Cr impurity in place of V in LiVO$_2$. Fig.~\ref{fig.d3impurities} shows the resulting textures. The four panels in this figure are directly comparable to Fig.~\ref{fig.impurities} of the main text -- the only difference being a d$^3$ impurity rather than a d$^1$ impurity. The textures are nearly identical, except for one additional bond at the d$^3$ impurity. This additional bond converts the loop-termination point of d$^1$ into a T-junction. Indeed, any d$^1$ texture can be converted into one with d$^3$ by the addition of one extra bond.

\end{document}